\documentclass[11pt]{revtex4}    
\usepackage{amssymb,epsf}
\usepackage{latexsym}

\begin{document}

\title{Rotating Solutions of Einstein-Maxwell-Dilaton Gravity with Unusual Asymptotics}

\author{A. Sheykhi  and  N. Riazi,\footnote{email address:
riazi@physics.susc.ac.ir}}

\address{Physics Department and Biruni Observatory, College of Sciences,
Shiraz University, Shiraz 71454, Iran}

\begin{abstract}
We study  electrically  charged, dilaton black holes, which
possess infinitesimal angular momentum in the presence of one or
two Liouville type potentials. These solutions are neither
asymptotically flat nor (anti)-de Sitter. Some properties of the
solutions are discussed.

\end{abstract}

\maketitle 
\section{Introduction}
Recently, non-asymptotically flat black hole spacetimes have been
of much interest in  the framework of AdS/CFT correspondence.
 Black hole spacetimes which are neither asymptotically flat nor dS/AdS
 have been found and investigated by
many authors. The first uncharged solutions were found by Mignemi
and Wiltshire  \cite{MW1}, Wiltshire  \cite{W} and Mignemi and
Wiltshire  \cite{MW2}. Global properties of the
Einstein-Maxwell-dilaton (EMd) gravity with a Liouville potential
were first obtained and discussed  by Poletti and Wiltshire
\cite{PW}. Static, electrically or magnetically charged,
non-asymptotically flat, non-dS/AdS black holes in various
dimensions were also found and discussed in \cite{CW}.
 The exact static and spherically symmetric solutions of
the electrically or magnetically charged dilaton black holes in
$n$ dimensions in the presence of one and two Liouville type
potentials and with unusual asymptotics (neither flat nor (anti)
de Sitter) were introduced by Chan, Horne and Mann\cite{CHM}.\\
The exact solutions mentioned above are all static. Recently,
magnetic, rotating solutions in four dimensional
Einstein-Maxwell-dilaton gravity with Liouville-type potential has
been constructed by Dehghani \cite{deh1}. These solutions are not
black holes, and present spacetimes with conic singularity.
Electrically charged rotating dilaton black strings were also
obtained and discussed in \cite{deh2}. Till now, charged rotating
dilaton black hole solutions for an arbitrary coupling constant
have not been constructed. Indeed,  exact solution for a rotating
black hole with a special dilaton coupling was derived using the
inverse scattering method \cite{BR}. For general dilaton coupling,
the properties of asymptotically flat charged dilaton black holes
only with infinitesimal  angular momentum \cite{shir} or small
charge \cite{CHL} have been investigated. Stationary rotating
black holes in $SU(2)$ Einstein-Yang-Mills theory, coupled to a
dilaton considered by Kleihaus \textit{et al.}\cite{kle}. These
black holes possess non-trivial non-Abelian electric and magnetic
fields outside their regular event horizon. Some classes of
solutions of non asymptotically flat, non AdS/dS charged dilaton
black holes with infinitesimal small angular momentum, and with
one Liouville type potential, were also discussed by Ghosh and
Mitra \cite{mitra}. In this paper, we present electrically
charged, dilaton black holes, for an arbitrary value of coupling
constant with infinitesimal angular momentum. We consider three
cases: without potential, with one Liouville type potential and
two Liouville type potential. These solutions are neither
asymptotically flat nor AdS/dS.
\\The organization of this paper is as follows:
After introducing the general equations of motion, we present and
discuss rotating dilaton black holes without potential. In section
\ref{4}, we present two classes of rotating  solutions with a
Liouville type potential and general dilaton coupling. In section
\ref{5}, we generalized these rotating solutions for the case of
two Liouville potentials. The last section is devoted to some
concluding remarks.

\section{Field Equations}
We consider the four-dimensional action in which gravity is
coupled to  dilaton and Maxwell  fields with an action
\begin{equation}\label{action1}
 S = \int d^4x\sqrt{-g}\left[{\cal R}
-2(\nabla{\phi})^2 - V(\phi) -
e^{-2\alpha\phi}{F^{\mu\nu}F_{\mu\nu}}\right]
\end{equation}
where ${\cal R}$ is the Ricci scalar curvature  and $\phi$ is the
dilaton field and $V(\phi)$ is a potential for $\phi$.
 The equations of motion can be
obtained by varying the action (\ref{action1}) with respect to the
gauge field $A_{\mu}$, the metric $g_{\mu\nu}$ and  the dilaton
field $\phi$ which yields the following field equations
\begin{equation}\label{FE1}
{\cal R}_{\mu\nu} = 2\partial_{\mu}\phi\partial_{\nu}\phi
+\frac{1}{2}g_{\mu\nu}V(\phi) +2e^{-2\alpha\phi}\left( F_{\mu\eta}
F_{\nu}^{\eta}
-\frac{1}{4}g_{\mu\nu}F_{\lambda\eta}F^{\lambda\eta}\right),
\end{equation}
\begin{equation}\label{FE2}
\partial_{\mu}\left(\sqrt{-g}e^{-2\alpha\phi} F^{\mu\nu}\right)=0,
\end{equation}
\begin{equation}\label{FE3}
\nabla^{\mu}\nabla_{\mu}\phi =\frac{1}{4}\frac{\partial
V}{\partial
\phi}-\frac{\alpha}{2}e^{-2\alpha\phi}F_{\lambda\eta}F^{\lambda\eta}.
\end{equation}
We wish to find rotating solutions of the above field equations.
For infinitesimal angular momentum, we can take the metric of the
form
\begin{equation}\label{metric}
ds^2 = - U(r)dt^2 + {dr^2\over U(r)} + R^2(r)\left(d\theta^2 +
\sin^2\theta d\phi^2 \right)-2a f(r)\sin^{2}{\theta}dt d{\phi}.
\end{equation}
Here, $a$ is an angular momentum parameter and $f(r)$ is a
function to be determined. Note that, in the particular case
$a=0$, this metric reduces to the static and spherically symmetric
case. For small $a$, we except to have solutions with $U(r)$ still
a function of $r$ alone. \\First of all, the $t$ component of the
Maxwell equation can be integrated immediately to give
\begin{equation}\label{Ftr}
 F^{rt}= \frac{q e^{2\alpha\phi}}{ R^2},
\end{equation}
where $q$, is the electric charge. In general, in the presence of
rotation, there is a vector potential
\begin{equation}\label{Aphi}
 A_{\phi}={a q h(r)\sin^2\theta}.
\end{equation}

 With the metric (\ref{metric}) and Maxwell fields (\ref{Ftr}) and
 (\ref{Aphi}), the field equations reduce to the following system of coupled
ordinary differential equations

\begin{equation}\label{ODE1}
\frac{1}{R^2}{d\over dr }\left(R^2 U {d\phi\over dr } \right)
=\frac{1}{4}\frac{d{V}}{d\phi}+\alpha
e^{2\alpha\phi}\frac{q^2}{R^4},
\end{equation}
\begin{equation}\label{ODE2}
{1\over R}{d^2R\over dr^2} + \left({d\varphi\over dr }\right)^2=0,
\end{equation}
\begin{equation}\label{ODE3}
{1\over R^2}{d\over dr }\left(U {dR^2\over dr } \right)=
\frac{2}{R^2}-V(\phi)- 2 e^{2\alpha\phi}\frac{q^2}{R^4}
\end{equation}
\begin{equation}\label{ODE4}
R^{4}\frac{d^2U}{dr^2}-2 R^{2}U ({\frac{dR}{dr}})^{2}-2 R^{3}U
({\frac{d^{2}R}{dr^{2}}})+ 2{R^2} - 4q^{2} e^{2\alpha\phi}=0,
\end{equation}
In addition, we have two coupled differential equations for
arbitrary functions $f(r)$ and $h(r)$.
\begin{equation}\label{ODE5}
R^{2}\frac{d^2f}{dr^2}-2f({\frac{dR}{dr}})^{2}
-2fR\frac{d^2R}{dr^2} -4q^{2}{\frac{dh}{dr}}=0,
\end{equation}
\begin{equation}\label{ODE6}
R^{2}\frac{d}{dr}(Ue^{-2\alpha\phi}\frac{dh}{dr})-
R^2\frac{d}{dr}(\frac{f}{R^2}) -2he^{-2\alpha\phi}=0,
\end{equation}
These equations  which arise from the presence of $A_\phi$, appear
only when $a\neq 0$, while the other equations were there also in
the static, spherically symmetric case. Static solutions of these
equations with unusual asymptotic were given  in \cite{CHM}.
\\To solve these equations, we make the ansatz
\begin{equation}\label{Ransatz}
R(r)=e^{\alpha \phi(r)},
\end{equation}
Using (\ref{Ransatz}) in equation (\ref{ODE2}), immediately gives
\begin{equation}\label{phi}
 \phi(r)=\frac{\alpha}{1+\alpha^2}\ln(br-c),
\end{equation}
where $b$ and $c$ are integration constants. For later
convenience, without loss of generality, we set $b=1$ and $c=0$.

\section{Solutions With $ V(\phi)=0$}
We begin by looking for the solutions without Liouville potential
$(V(\phi)=0)$. In this case equations (\ref{ODE1})-(\ref{ODE4}),
gives the following solution
\begin{equation}\label{U1}
U(r)= r^{2-2N}(1-\frac{2M}{Nr}),
\end{equation}
with $N={\frac{\alpha^2}{1+\alpha^2}}$ and M is the quasilocal
mass \cite{BY}. Note that the solution is ill defined for
$\alpha=0$. In order, that this solution  satisfy in all field
equations, we should have the following relation for the electric
charge:
\begin{equation}\label{q1}
q^2=\frac{1}{1+\alpha^2},
\end{equation}
There is an event horizon at $ r_{+}=\frac{2M}{N}$. In the limit
$\alpha^2\rightarrow \infty$, the electric charge vanishes and the
metric reduces to the Schwarzschild black hole.\\Here we are
interested  in finding rotating version of these solutions. For
infinitesimal rotation parameter $a$, we can get a solution of the
full set of equations with $U(r)$, $R(r)$ and $\phi(r)$ unaltered
and supplemented by the solutions of the new equations
(\ref{ODE5}) and (\ref{ODE6}) for two unknown functions $f(r)$ and
$h(r)$. To solve these equations we try the power relation
\begin{equation}\label{h}
h(r)=kr^m,
\end{equation}
where $k$ is a constant. Using this anzats,  we can distinguish
different solutions:\\
For $m=0$, one obtains the following solutions for $f(r)$:
\begin{eqnarray}\label{f1}
f_{1}(r)=\frac{2k}{4N-1}r^{1-2N},\\
f_{2}(r)=c r^{2N}+ f_{1}(r).
\end{eqnarray}
where $c$ is an integration constant. Note that these solutions
are ill defined for  $\alpha^{2}=\frac{1}{3}$ and in the case
$\alpha=1$, $f_{1}(r)$ becomes constant, while $f_{2}(r)=c r$.\\
In the particular case $k=0$, this solution involves a change of
the metric from the non-rotating form without any change of the
Maxwell field and follows from the general structure of the new
equations for $f(r)$ and $h(r)$. This may be surprising at first
sight because a rotation enters the metric without any rotation in
the charge; this is possible because the function $f(r)$ does not
obey conventional boundary conditions for large $r$ and in fact
increases with $r$.\\
In the large $\alpha$ limit,  from (\ref{q1}) we have $q^2=0$. For
$f(r)$, one obtains
\begin{equation}\label{f2}
 f(r)=-2kmMr^{m}+cr^2,
\end{equation}
with $c$ constant and $m=-1 , 2$. This is similar to the slowly
rotating Kerr black hole. The corresponding static case is the
Schwarzschild black
hole.\\
In addition, there are \textit{asymptotic} solutions  such as
\begin{equation}\label{f3}
 f(r)=r^{2N}\left(c+\frac{4kq^{2}}{\gamma}r^{\gamma}\right),
\end{equation}
with $\gamma=m+1-4N$ and $c$ is a constant. In this case, $\alpha$
is related to $m$  via
\begin{equation}\label{alpm}
 \alpha^{2}=\frac{m^2+m-6}{3m-m^2+2}.
\end{equation}
Note that this solution is ill defined for $\gamma=0$ or $
\alpha^2=\frac{m+1}{3-m}$.
\section{Solution With a Liouville type potential }\label{4}
 In this section, we consider the action (\ref{action1}) with a
Liouville type potential,
\begin{equation}\label{v1}
 V(\phi)=2\Lambda e^{2\beta\phi},
\end{equation}
where $\Lambda$ and $\beta$ are constants. In this case, equations
(\ref{ODE1})-(\ref{ODE4}) admit two classes of solutions.\\
i)  For the  first class of solutions, we obtain
\begin{equation}\label{U5}
U(r)= r^{2-2N}\left(1-\frac{2M}{Nr}+
\frac{\Lambda(1+\alpha^2)^2}{\alpha^2(1-3\alpha^2)}
r^{2(2N-1)}\right),
\end{equation}
 with $\beta=\frac{-1}{\alpha}$. In order, to satisfy this solution in all field
equations, the electric charge should be related to $\alpha$ via
 eqn. (\ref{q1}). Note that the solution is ill defined for
$\alpha^2=\frac{1}{3}$ and $\alpha=0$. In the limit
$\Lambda\rightarrow 0$ the solution reduces to that with
$V(\phi)=0$. On the other hand, when $\alpha^2\rightarrow \infty$
the solution becomes
\begin{equation}\label{SCH}
U(r)= 1-\frac{2M}{r}- \frac {\Lambda}{3}r^2,
\end{equation}
which is the Schwarzschild  dS/AdS  black hole, depending on the
sign of $\Lambda$. In order to investigate the causal structure of
the solution, we must investigate  the zeros of the metric
function $U(r)$. In fact, for $0<r <\infty$ the zeros of $U(r)$
are governed by the function
\begin{equation}\label{fr}
F(r) =1-\frac{2M}{Nr}+
\frac{\Lambda(1+\alpha^2)^2}{\alpha^2(1-3\alpha^2)} r^{2(2N-1)} .
\end{equation}
We investigate the function $g(r)=rF(r)$, for simplicity. The
cases with $\alpha^2>1/3$ and $\alpha^2<1/3$ should be considered
separately. We  should also consider the sign of the parameter
$\Lambda$ in each case. In the first case where $\alpha^2<1/3$ and
$\Lambda<0$ we may have one horizon since $\frac{dF}{dr}>0$. But
the more interesting case happens for $\Lambda>0$ where we obtain
only one local minimum at $r=r_{min}$ where
\begin{equation}\label{rmin}
r_{min}=\left(\frac{\alpha^{2}}{\Lambda(\alpha^{2}+1)}\right)^{\frac{1}{2}\frac{\alpha^{2}+1}{\alpha^{2}-1}}
\end{equation}\\
The function $F(r)$ possesses zeros if $F(r_{min})\leq 0$. There
are two zeros for $F(r_{min})< 0$ and one degenerate zero for
$F(r_{min})=0$ which corresponds to an extremal black hole. The
condition $F(r_{min})\leq 0$ gives
\begin{equation}\label{nn}
M\geq\frac{\alpha^{2}(2-3\alpha^{2})}{2(1+\alpha^{2})(1-3\alpha^{2})}
\left(\frac{\alpha^{2}}{\Lambda(1+\alpha^{2})}\right)
^{\frac{1}{2}\frac{\alpha^{2}+1}{\alpha^{2}-1}}
\end{equation}\\
In the second case  for $\Lambda<0$, the function $g(r)$ increases
monotonically. So we can conclude that  there is one point where
$F(r)=0$ which is the black hole horizon. For $\Lambda>0$ we find
local extremum for the function. The sign of
$\frac{d^{2}g(r)}{d^{2}r}$ determines weather we have local
maximum or minimum. For $\alpha^{2}>1$ we have local maximum and
$F(r_{max})$ should be positive in order to have any horizon. The
latter condition gives
\begin{equation}\label{nnn}
M\leq\frac{\alpha^{2}(2-3\alpha^{2})}{2(1+\alpha^{2})(1-3\alpha^{2})}
\left(\frac{\alpha^{2}}{\Lambda(1+\alpha^{2})}\right)
^{\frac{1}{2}\frac{\alpha^{2}+1}{\alpha^{2}-1}}
\end{equation}\\
If we have $\frac{1}{3}<\alpha^{2}<1$, then we would have local
minimum and in case of any horizon existing $F(r_{min})$ in eqn.
(\ref{rmin}) should be negative which implies eqn. (\ref{nn}). The
above considerations show that the solutions describe black holes
with two horizons or an extremal black hole hiding a singularity
at the origin $r=0$, when the mass satisfies (\ref{nn}) or
(\ref{nnn}). The radius of the inner and outer horizons can not be
expressed in a closed analytical form except for the extremal
case. The radius of the extremal solution coincides with $r_{min}$
\begin{equation}
r_{ext}=r_{min}=\left(\frac{\alpha^{2}}{\Lambda(\alpha^{2}+1)}\right)^{\frac{1}{2}\frac{\alpha^{2}+1}{\alpha^{2}-1}}=
\frac{2(1+\alpha^{2})(1-3\alpha^{2})}{\alpha^{2}(2-3\alpha^{2})}M
\end{equation}
Unfortunately, because of the nature of the exponents of $r$ in
(\ref{fr}), the event horizon determined by $F(r)=0$ can not be
expressed in a closed analytical form for arbitrary $\alpha$.\\In
order to obtain the rotating version of this solutions, we must
solve eqn.(\ref{ODE5}) and (\ref{ODE6}) for the two unknown
functions $f(r)$ and $h(r)$. Using the anzats (\ref{h}) we can
distinguish
different solutions:\\
For $m=0$, once again we have solution of the form (\ref{f1})
\begin{eqnarray}\label{f11}
f_{1}(r)=\frac{2k}{4N-1}r^{1-2N},\\
f_{2}(r)=c r^{2N}+ f_{1}(r).
\end{eqnarray}
In particular case $k=0$, we have unusual solution where there is
a change in the metric from the non-rotating form without any change in the Maxwell field. \\
In the large $\alpha$ limit,  we have a solution for $m=-1$. In
this case, from (\ref{q1}) we have $q^2=0$. For $f(r)$, one
obtains
\begin{equation}\label{f4}
 f(r)=\frac{2kM}{r}+\frac{\Lambda k}{3}r^2+cr^2,
\end{equation}
with $c$ constant. This is the form for slowly rotating Kerr
Ads/ds black hole. The corresponding static case is the
Schwarzschild  AdS/dS black hole which  present in (\ref{SCH}).
\\
In addition, for $\alpha^2<1$ there are \textit{asymptotic}
solutions such as
\begin{equation}\label{f3}
 f(r)=r^{2N}\left(c+\frac{4kq^{2}}{\gamma}r^{\gamma}\right),
\end{equation}
with $\gamma=m+1-4N$ and $c$  constant.\\
For $\alpha^2=1$, there exist an exact solutions such as
\begin{equation}\label{f4}
 f(r)=\frac{4kq^2}{m-1}r^m-4kmMr^{m-1}+cr,
\end{equation}
with $\Lambda=\frac{m^2-m-4}{2m(m-1)}$. Note that for $m\neq  2$
this solution exists only for $M=0$. For $m=0 , 1$, the solution
doesn't exist since $\Lambda$ diverges.\\

ii)For the second class of solutions, we obtain
\begin{equation}\label{U4}
U(r)= r^{2-2N}\left(1-2\Lambda -\frac{2M}{Nr}\right),
\end{equation}
 with $\beta=-\alpha$. For the electric charge one
obtains
\begin{equation}\label{q4}
q^2=\frac{1+\Lambda(\alpha^2-1)}{1+\alpha^2}.
\end{equation}
There is an event horizon at $ r_{+}=\frac{2 M}{N(1-2\Lambda)} $
which is regular only for $\Lambda < \frac{1}{2}$. In the limit
$\Lambda\rightarrow 0$ this solution reduces to that with
$V(\phi)=0$. Here we are interested in finding the rotating
version of this static solution, that is to say, in solving the
corresponding coupled equations for two unknown functions $f(r)$
and $h(r)$. By the anzats (\ref{h}) we can distinguish
different solutions:\\
For $m=0$, once again we have solution of the form (\ref{f1}). In
the large $\alpha$ limit, and $q=0$  we have also solutions as
long as
\begin{equation}\label{f6}
f(r)=-2kmMr^{m}+cr^2,
\end{equation}
with $c$ constant and $\Lambda=\frac{m^2-m-4}{2m(m-1)}$. Note that
for $m\neq  2,-1$ this solution exist only for $M=0$. For $m=0,1$,
the solution doesn't exist since $\Lambda$ diverges. For
\textit{asymptotic} solutions once  again we have solutions of the
form
\begin{equation}\label{f3}
 f(r)=r^{2N}\left(c+\frac{4kq^{2}}{\gamma}r^{\gamma}\right),
\end{equation}
with $\gamma=m+1-4N$ and $c$  constant. In this case, $\Lambda$ is
related to $m$ and $\alpha$ via the
\begin{equation}\label{Lambda}
\Lambda=\frac{(m^2+m-6)+\alpha^2(m^2-3m-2)}{2(m^2+m-2)+2\alpha^2(m^2-3m+2)}.
\end{equation}

 For $\alpha^2=1$, there exists the following exact solution
\begin{equation}\label{f4}
 f(r)=\frac{4kq^2}{m-1}r^m-4kmMr^{m-1}+cr,
\end{equation}
with $\Lambda=\frac{m^2-m-4}{2m(m-1)}$. Note that for $m\neq  2$
this solution exists only for $M=0$. For $m=0,1$, the solution
doesn't exist since $\Lambda$ diverges.\\
For $h(r)=r^2+\frac{2M(\alpha^2-1)}{\alpha^2}r$ and
$\Lambda=\frac{-\alpha^2}{2}$, one can also find solutions like
\begin{equation}\label{f4}
 f(r)=2(1+\alpha^2)\left(\frac{\alpha^2-2}{\alpha^2-3}r^{3-2N}
 +M\frac{\alpha^2-2}{\alpha^2}r^{2-2N}+2M^2\frac{1-\alpha^2}{\alpha^4}r^{1-2N}+cr^{2N}
 \right),
\end{equation}
with $N={\frac{\alpha^2}{1+\alpha^2}}$ and $c$  constant.
\section{Solutions with a general  coupling  parameter  and  two  Liouville
potentials}\label{5}

In this section, we present rotating solutions to the EMd gravity
equations with infinitesimal rotation parameter  and dilaton
potential
\begin{equation}\label{v2}
V(\phi) = 2\Lambda_{1} e^{2\beta_{1}\phi} +2 \Lambda_{2}
e^{2\beta_{2}\phi},
\end{equation}
 where $\Lambda_{1}$ and
$\Lambda_{2}$ are constants. This generalizes further the
potential (\ref{v1}). If $\beta_{1}=\beta_{2}$, then (\ref{v2})
reduces to (\ref{v1}), so we will not repeat these solutions.
Requiring $\beta_{1}\neq \beta_{2}$, from equations
(\ref{ODE1})-(\ref{ODE4}), one obtains
\begin{equation}\label{U6}
U(r)= r^{2-2N}\left(1-2\Lambda_{1}-\frac{2M}{Nr}+
\frac{\Lambda_{2}(1+\alpha^2)^2}{\alpha^2(1-3\alpha^2)}
r^{2(2N-1)} \right),
\end{equation}
with $N={\frac{\alpha^2}{1+\alpha^2}}$ and M is the quasilocal
mass. In order to fully satisfy the system of equations, the
$\beta_{1}$ and $\beta_{2}$ parameters must satisfy
$\beta_{1}=\frac{1}{\beta_{2}}=-\alpha$. For the electric charge,
one obtains
\begin{equation}\label{q6}
q^2 = \frac{1+\Lambda_{1}(\alpha^2-1)}{1+\alpha^2},
\end{equation}
Obviously, another solution with the same spacetime metric is
generated via the discrete transformation
$\beta_{1}\longleftrightarrow
\beta_{2}$ and $\Lambda_{1}\longleftrightarrow \Lambda_{2}$.\\
Note that in the particular case $\Lambda_{2}=0$, this solution
reduces to (\ref{U4}) and  when $\Lambda_{1}=0$, it reduces to
(\ref{U5}). In order to investigate the causal structure of the
solution and subsequently find the horizons (similar to what was
done in the previous section) we find the zeros of the function
\begin{equation}
F(r)=-\frac{2M}{N}+(1-2\Lambda_{1})r+\frac{\Lambda_{2}(1+\alpha^{2})}{\alpha^{2}(1-3\alpha^{2})}r^{4N-1},
\end{equation}
\par
We consider the cases $\alpha^{2}>1/3$ and $\alpha^{2}<1/3$,
separately. For the first case, we certainly have extremum if
$\Lambda_1>1/2(\Lambda_1<1/2)$ or $\Lambda_2<0(\Lambda_2>0)$. The
sign of the second derivative will show whether we have local
minimum or maximum. Here, for $1/3<\alpha^2<1(\alpha^2>1)$ and
$\Lambda_2>0(\Lambda_2<0)$ the function $f(r)$ would have local
minimum and in opposite, for $1/3<\alpha^2<1(\alpha^2>1)$ and
$\Lambda_2<0(\Lambda_2>0)$ the function will have local maximum at
\begin{equation}\label{rmax}
r_{min(max)}=\left(\frac{(1-2\Lambda_1)\alpha^2}{\Lambda_2(\alpha^2+1)}\right)
^{\frac{1}{2}\frac{\alpha^2+1}{\alpha^2-1}}.
\end{equation}
The value of the function $f(r)$ at its extremum is
\begin{equation}\label{frmax}
f(r_{extr})=-\frac{2M}{N}+\frac{(1-2\Lambda_1)(2-3\alpha^2)}{1-3\alpha^2}
\left(\frac{(1-2\Lambda_1)\alpha^2}{\Lambda_2(\alpha^2+1)}\right)^{\frac{1}{2}\frac{\alpha^2+1}{\alpha^2-1}}
\end{equation}
In order to have any horizon, $F(r_{min})(F(r_{max}))$ should be
larger(less) than or equal to zero in order to possess  any local
extremum and subsequently to have any horizon for the black hole.
The case $F(r_{min})(F(r_{max}))=0$ corresponds to an extremal
black hole. The condition $F(r_{min})<0$ gives
\begin{equation}\label{namosavi}
M\geq\frac{\alpha^2(2-3\alpha^2)(1-2\Lambda_1)}{2(1+\alpha^2)(1-3\alpha^2)}
\left(\frac{(1-2\Lambda_1)\alpha^2}{\Lambda_2(\alpha^2+1)}\right)^{\frac{1}{2}\frac{\alpha^2+1}{\alpha^2-1}}
\end{equation}
and we obtain the following inequality for the condition
$f(r_{max})>0$
\begin{equation}\label{namosavi}
M\leq\frac{\alpha^2(2-3\alpha^2)(1-2\Lambda_1)}{2(1+\alpha^2)(1-3\alpha^2)}
\left(\frac{(1-2\Lambda_1)\alpha^2}{\Lambda_2(\alpha^2+1)}\right)^{\frac{1}{2}\frac{\alpha^2+1}{\alpha^2-1}}.
\end{equation}
\par
For the second case where $\alpha^2<1/3$ the function $F(r)$
possesses local minimum for $\Lambda_2<0$ and local maximum for
$\Lambda_2>0$. In this case, the function diverges both at $r=0$
and at infinity. The local minimum (maximum) happens at
(\ref{rmax}) and the value of the function $F(r_{min})$ is given
by (\ref{frmax}). Since in this case we have both a local minimum
and maximum, condition (\ref{namosavi}), should hold for both
cases.
\par
We see that in both cases we obtain horizons for any given value
of the parameter $\alpha$. Here we express the  radius of the
extremal solution as in the preceding section
\begin{equation}
r_{ext}=\left(\frac{(1-2\Lambda_1)\alpha^2}{\Lambda_2(\alpha^2+1)}\right)
^{\frac{1}{2}\frac{\alpha^2+1}{\alpha^2-1}}=\frac{2(1+\alpha^2)(1-3\alpha^2)}{\alpha^2(2-3\alpha^2)(1-2\Lambda_1)}M
\end{equation}
Here, we are interested in finding the rotating version of this
static solution, i.e. solving the corresponding coupled equations
for two unknown functions $f(r)$ and $h(r)$. By the anzats
(\ref{h}) we can distinguish
different solutions:\\
For $m=0$, ones again we have solution of the form (\ref{f1}). In
particular case $k=0$, we have unusual solution where there is a
change in the metric from the non-rotating form without any change
in the Maxwell field also exist as before.\\
 In the large $\alpha$ limit, and $q=0$, one obtains
\begin{equation}\label{f8}
 f(r)=\frac{-km\Lambda_{2}}{3}r^{m+3}-2kmMr^{m}+cr^2.
\end{equation}
with $m=-1,-4$  and $\Lambda_{1}=\frac{m^2-m-4}{2m(m-1)}$. Note
that for $m=-4$ this solution exists only for $M=0$. For $m=-1$,
the solution  exists for $M$ non zero. For \textit{asymptotic}
solutions once again we have solutions of the form
\begin{equation}\label{f3}
 f(r)=r^{2N}\left(c+\frac{4kq^{2}}{\gamma}r^{\gamma}\right).
\end{equation}
with $\gamma=m+1-4N$ and $c$ is a constant. In this case $\Lambda$
is related to $m$ and $\alpha$ via
\begin{equation}\label{Lambda1}
\Lambda_{1}=\frac{(m^2+m-6)+\alpha^2(m^2-3m-2)}{2(m^2+m-2)+2\alpha^2(m^2-3m+2)}.
\end{equation}
For $\alpha^2=1$, there exists an exact solution in the form
\begin{equation}\label{f4}
 f(r)=\frac{4kq^2}{m-1}r^m-4kmMr^{m-1}+cr.
\end{equation}
In this case, $\Lambda_{2}$ is related to $m$ and $\Lambda_{1}$
via
\begin{equation}\label{Lambda2}
\Lambda_{2}=\frac{(m^2-m-4)-2\Lambda_{1}(m^2-m)}{2m(m-1)}.
\end{equation}
It is notable  that for $m\neq  2$ this solution exists only for
$M=0$. For $m=0,1$, the solution
doesn't exist since $\Lambda$ diverges.\\

\section{Conclusion}
In summary, we considered exact, electrically charged, static and
spherically symmetric black hole solutions to four dimensional
 Einstein-Maxwell-dilaton gravity without  potential or with  one or two Liouville
 type  potentials. These black holes have unusual asymptotics.
 They are neither asymptotically flat nor asymptotically (anti-) de
 Sitter.\\
We have added an infinitesimal rotation represented by the
parameter $a$. In this case we need only to know a few extra
components of the gauge field and the metric. These are
$A_{\phi}$, $A_{t}$ and $g_{t\phi}$ which are of order $a$. For
small angular momentum, the field equations led to the coupled
differential equations satisfied by two unknown functions $f(r)$
and $h(r)$, for which we  presented several classes of solutions.

\section*{Acknowledgements}
N. Riazi acknowledges the support of Shiraz University.


\end{document}